# Performance-aware placement and chaining scheme for Virtualized Network Functions: A Particle Swarm Optimization Approach

Samane Asgari[1] . Shahram jamali[2] . Reza Fotohi[3] . Mahdi Nooshyar[4]

**Abstract:** Network functions virtualization (NFV) is a new concept that has received the attention of both researchers and network providers. NFV decouples network functions from specialized hardware devices and virtualizes these network functions as software instances called virtualized network functions (VNFs). NFV leads to various benefits, including more flexibility, high resource utilization, and easy upgrades and maintenances. Despite recent works in this field, placement and chaining of VNFs need more attention. More specifically, some of the existing works have considered only the placement of VNFs and ignored the chaining part. So, they have not provided an integrated view of host or bandwidth resources and propagation delay of paths. In this paper, we solve the VNF placement and chaining problem as an optimization problem based on the particle swarm optimization (PSO) algorithm. Our goal is to minimize the required number of used servers, the average propagation delay of paths, and the average utilization of links while meeting network demands and constraints. Based on the obtained results, the algorithm proposed in this study can find feasible and high-quality solutions.

**Keywords** Network functions virtualization (NFV) . Virtualized Network Functions (VNFs) . Placement and chaining . Particle swarm optimization

✉ Samane Asgari
  Samane.asgari@student.uma.ac.ir

✉ Shahram jamali*
  Jamali@uma.ac.ir

✉ Reza Fotohi
  R_fotohi@sbu.ac.ir; Fotohi.reza@gmail.com

✉ Mahdi Nooshyar
  Nooshyar@uma.ac.ir

[1,2,4] Computer Engineering Department, University of Mohaghegh Ardabili, Ardabil, Iran
[3] Faculty of Computer Science and Engineering, Shahid Beheshti University, G. C. Evin, Tehran, Iran

# 1 Introduction

Today, intermediate boxes (or network functions (NF)) have become a pervasive element in networks. Firewalls, proxies, network address translators (NAT), and intrusion detection systems (IDS) are examples of network functions. Today's organizations include a large number of hardware intermediate boxes to meet the diverse demands of multiple streams across the network. These middle boxes are physically located at fixed locations in the network. Every traffic flow needs a special service that must be routed to an intermediate box of that service provider. A stream may not necessarily require just one network function. Instead, it may require navigating an ordered set of network functions, a phenomenon called service function chain (SFC) also called policy chain [1]. Service function chains may differ in the number and order of functions to navigate through streams.

Despite the growing popularity of middle boxes, they have become more of a one-time solution for network operators because:

- The processing capacity of a hardware middleware is fixed; such inflexibility limits their ability to operate efficiently and efficiently. For example, in times of low traffic, the middle box will run at full capacity, and therefore not fully utilized. On the other hand, in cases where there is a lot of traffic, it is possible for the middle boxes to serve more than their fixed capacities, which causes congestion in the network. To address these traffic waves, network operators tend to provide more hardware in the number and type of network functions purchased [2].
- Each function is located in a fixed physical location within the network. A request for a specific service chain may force traffic to go back and forth to navigate the network, causing it to take up unnecessary bandwidth from some links.

Intermediate boxes have high core costs (CAPEX) and operating costs (OPEX). They are expensive to buy, maintain and execute. In addition, their hardware life cycle is short, and rapid advances in technology, the addition of new services and agents, or the replacement of existing hardware require more procurement of hardware network functions, which is a tedious and tedious process [2].

Network Functionalization (NFV) was established in October 2012 by the Industrial Specification Group (ISG) under the auspices of the European Telecommunications Standards Institute (ETSI) and has grown significantly since then. Prior to NFV, software-based networking, or SDN, was introduced, which suggests that SDN was the premise for NFV.

NFV is a promising solution that solves the above problems. This process turns intermediate boxes into instances of simple software referred to as virtual network functions (VNFs) [3], thus eliminating the dependency between the network function and its hardware, and in sharing physical hardware between multiple VNFs. The shape of virtual machines or servers will work. In other words, virtualization of network functions is a process for designing, deploying and managing network infrastructure. Figure 1, shows a picture of network virtualization and its functions.

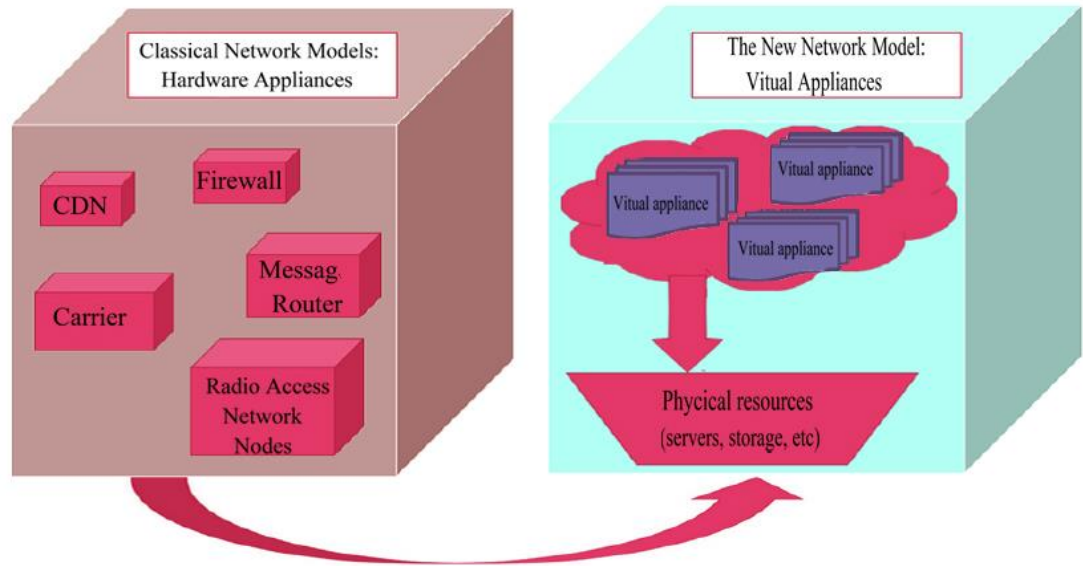

**Fig. 1** Virtualization of networks and their functions [1].

Optimal use of resources, more flexibility, easy updating, maintenance costs, reduction of capital costs (CAPEX) and operating costs (OPEX) of NFV are the most important benefits of NFV [1] [2]. The deployment and regulation of VNFs still requires further research and development in today's world [1]. The challenge of efficient placement of service chains requires more attention. A service chain is a sequence of VNFs that should be met in a specific order. The placement of a service chain consists of the allocation of servers (for hosting VNFs) and steering traffic flow from one VNF to the next VNF in the chain. Figure 2, shows a service function chain. Service function chains can consist of one or more different types of VNFs, as shown in Figure 2. As you can see, this service chain starts with a firewall, then continues with an intrusion detection system, and ends with a network address translator. Requests received at the beginning of each service function chain are processed in the First Input (FIFO) method.

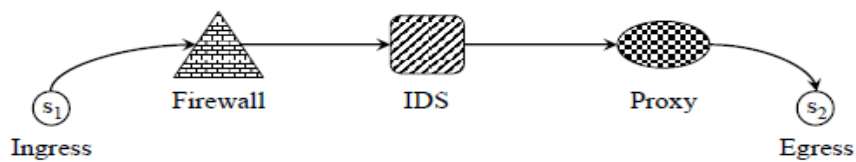

**Fig. 2** Service function chain.

An example of a VNF chain consisting of 5 nodes - which are considered as network servers to host VNFs - and meet two demands of the service chain. This example is shown in Figure 3. VNF has three different types as follows:

F1 (e.g. a firewall),

F2 (e.g. a gateway), and

F3 (e.g. a load balancer).

A path refers to the sequence of servers that specific traffic should pass through a particular service chain.

Fig. 3(a) represents the first service chain (SC1), at source node A and destination node D. It consists of two VNFs (F1 and F2). First, the function F1 should be met followed by function F2 and the second service chain (SC2) at source node A and destination node E. It consists of two VNFs (F1 and F3), which function F1 being met first followed by function F3.

Fig. 3(c) shows the placement of these two service chains. For the first service chain, the physical path A→ C → D is chosen, where node C hosts F1 and F2. For the second service chain, the physical path A→ C→ B → E is selected, where node B hosts F3. In this allocation, node C hosts an instance of F1 that is used by two demands to minimize the number of used resources.

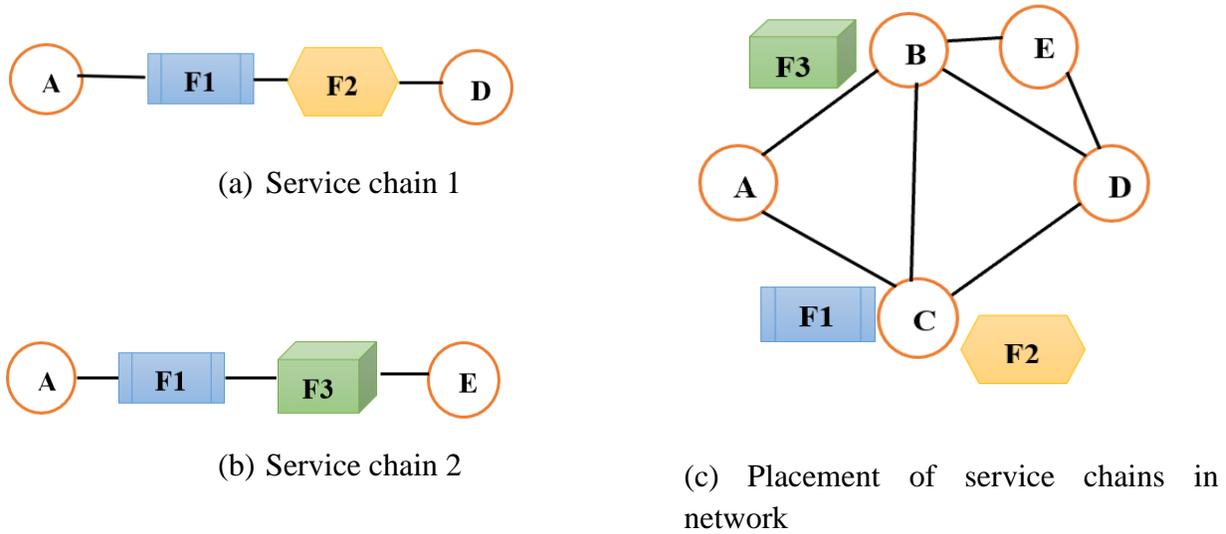

(a) Service chain 1

(b) Service chain 2

(c) Placement of service chains in network

**Fig. 3** An example of VNF placement.

Most of the existing works on placement and chaining of VNFs only consider a part of the problem. In other words, they optimize either host or bandwidth resources and they do not consider a complete scheme of computation, storage, and network optimization [3]. The authors of [4] consider only the bandwidth used for routing traffic through the selected paths while the authors of [5] consider the consumed bandwidth and compute the results within reasonable runtime with negligible performance effects.]. In the reference [6] only the criteria of the number of servers used to host VNFs are considered.

The focus of this paper is on VNF placement and chaining to use links, minimize the number of host nodes, and average propagation delay of selected paths. To solve this problem, we used PSO optimizer. PSO is a meta-analysis method used to obtain the best solution by updating the position and velocity of particles in the search space. This update is based on the particle size value [8]. This algorithm was chosen because of its simple implementation, quick convergence simple concept and calculation, and a limited number of parameters of PSO [9].

The remainder of this paper is organized as follows. Section 2 presents related works. Section 3 introduces the VNF placement and chaining strategies. Section 4 describes the proposed algorithm and formulates the problem. Performance evaluation results are reported in section 5. Finally, section 6 discusses the conclusions.

## 2  Related works

An example of the planning and deployment of physical intermediate boxes that show the placement and chain of network performance before the start of NFV [10]. However, today, it is one of the most important issues and many studies have been done in this field. In this section, we review some of the research works related to virtualized network function placement and chaining problems.

Marcelo Louiseley et al. [5] propose an approach to chain and location VNF with the aim of minimizing the allocation of required resources, while meeting network flow requirements and constraints. Their approach is to combine correct linear programming with an exploratory method (variable neighborhood search). Their approach is that they first compute an initial solution (or configuration) for the optimization problem, then select a subset of the network presence points iteratively, and the variables in the special list are optimized, while Other variables remain unchanged (or fixed, hence the name fixed and optimized).

Maroon Machteri et al. [6] discuss the aspect of locating service chains by finding the best locations and hosts for VNFs and directing traffic across network functions, while respecting user needs and maximizing provider benefits. They propose a new approach based on application components and infrastructure graphs for locating physical and virtual network chains in cloud environments and networks. The evaluation results show that their proposed algorithm reduces complexity and can be extended to thousands of nodes and links, as well as outperforms the relevant work in the provider's interests and acceptance rate.

Qu et al. [22] have proposed a MILP optimization model and an exploratory algorithm (Greedy-k-NFV) for VNF location and traffic routing that maximizes reliability and minimizes end-to-end service chain delays (SFCs). Their simulation results show that their heuristic algorithm performs better than the existing tasks in terms of reliability of 7.4% to 14.8% and in terms of average end-to-end latency of about 18.5%.

Sama Sola et al. [34] propose locating and chaining energy efficient VNFs across active NFV infrastructures. Their proposed approach is an extension of the Monte Carlo Tree Search Method (MCTS). Their proposed algorithm is called the Energy Efficient Search Tree Chain Location Algorithm (EE-TCA), which minimizes the energy consumption of the hardware (physical resources) and software (virtual resources) involved in hosting and optimizing the VNF chain. Their proposed algorithm performs the following steps:

1) initializing the tree,

2) selecting nodes to explore during tree orientation,

3) generating a new subdirectory,

4) optimizing node communication, and finally

5) selecting the best solution (saving more power).

The authors used a real cloud platform and extensive simulations for the proposed decision tree algorithm to evaluate the performance and ability to adapt to the problem size. This proposed algorithm shows the results of scalability simulations as well as a significant reduction in the energy consumption of the proposed solution compared to the dependent tasks.

An accurate and remarkable VNF locating solution called VNF-P is presented in [35], in which Mons et al. Used integer linear programming in a hybrid NFV environment that contained specific, all-purpose hardware. VNF-P, like much work in this area, aims to minimize the number of physical devices used to save substantial costs for a network provider. This has been evaluated (due to the scalability constraints of precise solutions) in small-scale networks with varying traffic loads, and provides results whose algorithms expire in 16 seconds or less to respond quickly to changes in traffic demands such as morning and night. Create the peak of traffic fluctuations in the evening.

Ahwar et al. [36] have defined the problem of locating and chaining VNFs as follows: Consider a chain of VNFs that must be placed on the NFV infrastructure provided for traffic flows requested from different sources and destinations. They hypothesized that in order to minimize the total cost, four criteria should be considered:

(1) how to find the right number of VNF specimens,

(2) how to find suitable locations for VNF specimens, and

(3) how to properly chain VNF specimens. And

(4) how the requested traffic is allocated to VNF chains.

They used integer linear programming to solve the problem of locating and chaining VNFs to minimize the overall cost. They used MATLAB software and AMPL-Gurobi version 1.5.6 to implement integer linear programming.

Mohammad Khan et al. [37] formulated the problem of locating network functions as a mixed integer linear program (MILP). This formulation not only determines the location of services and routing streams, but also seeks to minimize resource utilization. They have developed an exploration to solve this problem, which enables them to support large examples of the problem and solve the problem of incoming streams without affecting existing streams. Their approach shows that dividing a large problem into small parts can produce a result close to the optimal result.

Diye et al. [38] believe that locating and chaining VNFs affects the quality of service (e.g., latency) of a value-added service and the cost of providing the network provider. This is modeled as an optimization problem that aims to find the optimal number and location as well as efficiently chain VNF samples so that the cost of the network provider is minimized and the quality of service is met. The authors of this article consider the cost to include the cost of licenses, calculations, and the cost of communications. The number of VNF sites and samples used is calculated. Computing costs include the cost of running VNFs on servers and the cost of communications as the sum of the bandwidth used by the chains in the network. They have three main limitations to their heuristic algorithm: satisfying service quality while avoiding server overload and VNF. They designed their heuristic algorithm based on Google PageRank, the leading search engine algorithm. Their heuristic algorithm can be described as a kind of centralized vector method. PageRank also works well on free-scale networks and is therefore widely used in many fields.

Etkora et al. [40] considered the dynamic locating aspect of network virtual functions. They used genetic exploration algorithms to solve this problem. The particle in their path is a location. In their proposed algorithm, they consider minimizing propagation and packing latency and the number of cores used in physical machines (servers). Their results show that their proposed algorithm has a small probability of failure to find possible solutions, that is, it can track dynamic changes and, compared to other tasks, the location delay in updates is less.

Hawilo et al. [41] proposed a new MILP optimization model and a heuristic algorithm (BACON) for locating VNFs and their associated SFCs on a small-scale network with 30 servers and a large-scale network with 300 servers. Their purpose is to minimize intra and end-to-end service chain (SFC) delays. Their approach also increases the reliability and quality of service. Among the limitations that have been considered are capacity, network latency, accessibility, and so on. Their simulation results show that both of their proposed models perform better than the methods compared to them and have less time complexity.

Pei et al. [42] first solved the problem of selecting and chaining VNFs as a proper binary programming (BIP) model with the aim of minimizing end-to-end latency. Then they introduced a new two-phase algorithm based on deep learning. Their simulation results show that their two-phase algorithm is more efficient in terms of request acceptance rate, throughput, end-to-end delay.

## 3  Placement and chaining of VNFs strategies

In [10], it was proved that the VNF placement and chaining problem belongs to the complexity class NP-Complete. In this regard, the optimization algorithms to solve this problem can be divided into three categories: exact solutions, heuristic solutions, and meta-heuristic solutions.

Exact solutions usually use linear programming and can achieve the best result in a mathematical model. LP problems can be solved by software programs called solvers (i.e., CPLEX, GLPK, XPRESS, or IPOPT). Previous works [6], [12], [11], [4], [13], and [14] solve the problem of VNF placement and chaining by proposing an LP or ILP or MILP formulation.

Heuristic solutions are used to reduce the execution time to solve the placement and chaining of the VNFs problem. These solutions do not necessarily reach the best placement and chaining of VNFs such that they can be stuck in a local optimum and they depend on the problem [15]. However, they can achieve a good solution with a short execution time [16]. In [5], [17], [18], [7], [19], [20], [1], [10], [21], and [22], the authors used heuristic solution for this purpose.

Meta-heuristic algorithms, which are inspired by the process of natural selection, can find near-optimal solutions. These algorithms improve solutions by escaping from local optimum and their execution time is short. Since meta-heuristic algorithms are problem-independent, they can be used for so many problems [15].

Some of the popular meta-heuristic algorithms are genetic algorithm, PSO, and Ant Colony. There have only been a few studies to use a meta-heuristic approach for VNF placement and chaining [3], [24], and [25].

The authors of [25] and [24] only consider the VNF placement problem and propose their algorithm based on GA.

In another work [3], the authors considered the service chain placement problem, which requires selecting servers (to host VNFs) and paths to route traffic flow from one VNF to the next VNF in the chain. They propose their algorithm based on a GA and compare their algorithm with the ILP approach. They implement the ILP formulation in CPLEX. Their result shows that for small networks and few numbers of VNFs, ILP took 12.5 seconds while the genetic approach took 0.0034 seconds to find the optimal solution. Nevertheless, for a few VNFs in a large-scale network, the ILP can take hours and it may be stopped without finishing correctly but GA can find the optimal solution in a matter of seconds. Moreover, their result shows that solution quality (i.e., objective value, number of used servers, links, and link congestion) of GA is similar to ILP formulation.

# 4 Problem formulation and the proposed algorithm

In this paper, the chain problem and VNF placement are categorized into two main steps as follows:
Step 1, to direct the flow of traffic from one VNF to the next VNF in the chain, which involves selecting routes.
Step 2 Select the servers hosted by VNFs.

## 4.1 Problem formulation

We consider a network with N physical servers and L links. All the physical servers have equal capacity, which is denoted by $S_n$, and the physical server can host several VNFs. Similar to [23], a physical server is considered to use, if and only if it hosts at least one VNF; otherwise, it is considered as an idle or unused state. The set of used servers is denoted by T. A path (p) refers to the sequence of servers that specific traffic should pass through in a particular service chain. All links in the network have the equal capacity, which is denoted by $B_l$. The number of VNFs in the network is denoted by F and their type is denoted by M. When a VNF (f) is placed on a server (n), it occupies its necessary capacity of server capacity. Let RS be the set of demands in the network, each demand must determine requested VNFs, source server, and destination server [26]. For example, one demand is presented as:

$$RS1: f3 \rightarrow f2 \rightarrow f1 \text{ (source: 10; destination: 2)}$$

Each demand in the network has a special sequence of virtual functions to be met. It means that a virtual function (f) must not be met before its previous function in the chain. Now that the VNF placement and chaining problem is introduced, let us minimize the number of used servers *(T)*, the average utilization of links *(U)*, and the average propagation delay of selected paths ($\widehat{dp}$). It results in the servers that already have been used can be utilized efficiently and the traffic will not traverse long paths. The average propagation delay of all paths ($\widehat{dp}$) is defined as:

$$\widehat{dp} = \left( \frac{\sum_{p=1}^{P} d_p}{P} \right) \tag{1}$$

where $d_p$ is the propagation delay of path p and P is the number of all selected paths in the network. The objective function is normalized as follows:

$$Minimize\left( w_1 \frac{1}{N} * T + w_2 * U + w_3 * \frac{\widehat{dp}}{dp \max} \right) \tag{2}$$

where $w_1$, $w_2$, and $w_3$ are weighting factors by which the operators can determine the tradeoffs between these factors

N is the total number of physical servers, T is the total number of servers used, U is the average utilization of links, (dp) is the average propagation delay of paths, and dpmax is the maximum average propagation delay of all paths.

We considered the following constraints for the network. For each server used, the total capacity consumed by all the VNF instances assigned to that server does not exceed its capacity (3):

$$\sum_{f=1}^{F} I_f^n * C_f \leq S_n \qquad \forall n \in N \qquad (3)$$

In constraint (4), we ensure that for each link used, the total bandwidth consumed by all the VNF instances does not exceed the bandwidth of that link.

$$\sum_{p=1}^{P} \sum_{f=1}^{F} W_f^p * X_l^p * BW_f \leq B_l \qquad \forall l \in L \qquad (4)$$

In constraint (5), we ensure that the propagation delay of each path does not exceed the maximum average propagation delay of all selected paths.

$$\sum_{l=1}^{L} X_l^p * de_l \leq dp \max \qquad \forall p \in P \qquad (5)$$

In constraint (6), we guarantee that a server is active only when it hosts at least one VNF instance.

$$\sum_{p=1}^{P} \sum_{f=1}^{F} W_f^p * Q_n^p * I_f^n - G_n \geq 0 \qquad \forall n \in N \qquad (6)$$

Table I shows the variables that are used to formulate the optimization problem.

### 4.2 Particle Swarm Optimization

Dr. James Kennedy and Dr. Russel Eberhat in 1995 have introduced a population-based optimization method, called PSO. PSO is inspired by animal behavior, such as bird flocking and fish schooling [27]. In PSO, individuals are called particles and the population size is equal to the number of particles. Also, in this algorithm, each particle represents a solution and has a position (i.e., $X_i$) and velocity (i.e., $V_i$), which are updated in each iteration. The particle community optimization algorithm uses two basic branches of learning: social sciences and computer science. In addition, the PSO algorithm uses the theory of collective intelligence, which is a system for displaying cooperative and group activities/guiding non-knowledge agents that interact locally with their environment and create logical global performance patterns [28].

**Table 1** list of variables and their definition

| Variables | Definition |
|---|---|
| $N$ | Number of physical servers, indexed by $n=1,2,...,N$ |
| $L$ | Number of links, indexed by $l=1,2,...,L$ |
| $P$ | Number of selected paths, indexed by $p=1,2,...,P$ |
| $F$ | Number of VNFs, indexed by $f=1,2,...,F$ |
| $M$ | Types of functions |
| $T$ | Total number of used servers |
| $U$ | Average utilization of links |
| $dp\max$ | The maximum average propagation delay of all paths |
| $d\hat{p}$ | The average propagation delay of all paths |
| $d_p$ | the propagation delay of path p |
| $C_f$ | Required processing capacity of network function f |
| $BW_f$ | The required bandwidth capacity of network function f |
| $S_n$ | Processing capacity of server n |
| $B_l$ | The bandwidth capacity of link l |
| $G_n \in \{0,1\}$ | $G_n=1$ if server n is used, otherwise $G_n=0$ |
| $W_f^p \in \{0,1\}$ | $W_f^p=1$ if VNF f routes traffic on path p, otherwise $W_f^p=0$ |
| $X_l^p \in \{0,1\}$ | $X_l^p=1$ if link l is used in path p, otherwise $X_l^p=0$ |
| $I_f^n \in \{0,1\}$ | $I_f^n=1$ if VNF f is hosted on server n, otherwise $I_f^n=0$ |
| $Q_n^p \in \{0,1\}$ | $Q_n^p=1$ if server n is used in path p, otherwise $Q_n^p=0$ |
| $w_1, w_2, w_3$ | Weighting factors |

$$V_i(t+1) = \left(w*V_i(t) + c_1 r_1 *(p_i - X_i(t)) + c_2 r_2 *(p_g - X_i(t))\right) \qquad (7)$$

$$X_i(t+1) = \left(X_i(t) + V_i(t+1)\right) \qquad (8)$$

The two variables *r1* and *r2* are random variables whose value is between [0,1]. To control the effect of the current velocity on the next velocity, the variable *w*, which is the weight variable of inertia, is used. Acceleration factors were also determined by two variables *c1* and *c2*. The best position that all particles have experienced is set in the variable *pg*. pi is the best position that particle *i* has experienced, and the fitness function is used to evaluate particles [30].

> 1) Create initial population based on random speed and position
>
> 2) Calculate the amount of fitness function for each particle (Equation 1).
>
> 3) If the value of the fitness function is less than the value set for pi, set the current value as the new pi and the best value for pi as pg.
>
> 4) Use formulas 7 and 8 to calculate the particle velocity and update the position of the particles.
>
> 5) Repeat from step 2 until reaching the maximum iteration/minimum error.

In this article, each particle represents a full solution, which includes servers to host VNFs and paths assignment for serving all the service chains in the network. It means that, if the population size (i.e., number of particles) is equal to m, there are m different possible configuration states for the network that each of them determines selected servers to host VNFs and selected paths to meet all the demands.

Our algorithm uses topology data, demands, set of VNF types, and their capacity and server capacity as inputs. In the first step, our algorithm generates the initial population. It selects the initial assignment of servers to host network functions and paths for each demand. We used a random algorithm as the initial solution. In this way, the servers and paths are selected randomly and particle positions and velocities are randomly initialized.

In the second step, the fitness function (equation 1) for the initial population is measured to see how good a full solution is. Our fitness function takes into account the number of used servers, links capacity, and the propagation delay of selected paths. In the fitness function, weights of parameters ($w_1$, $w_2$, and $w_3$) depend on the preferences. It means that more priority can be given to server utilization or link utilization or average propagation delay of paths. In our implementation, more priority is given to server utilization. The best solutions are those that return the smallest value.

In the third step, for each particle, the particle's fitness function is compared with its best value, and if it is smaller than its best value, it will be set as the new best value and the particle's position will be stored. Then, the smallest value of the fitness function of all particles will be set as the global best value.

In the fourth step, the velocity and position of all particles are updated according to equations (6) and (7). This process is iterated until the termination criterion is reached. In our implementation, the maximum number of iterations is used as the termination criterion. In the final iteration, the solution that gives the smallest fitness function value is selected as the best solution to the problem.

## 5 Performance evaluation

In this section, we describe the experimental settings used in this article. In this experiment, three networks are used in different situations, where the networks do not already have a specific structure for VNF placement and chain problem. The proposed method is compared with the Random algorithm under different parameters because our design and purpose are different from previous articles on VNF and chain placement. Different scenarios have been used to simulate and evaluate the efficiency of the proposed algorithm. In each scenario, the value of some parameter's changes, in which case, their value is expressed for that particular scenario. In this section, the effect of number of requests on average release delay, request acceptance rate and the effect of service chain growth on execution time, average route release delay and the effect of number of servers on execution time are investigated and finally the proposed algorithm with random approach Is compared.

## 5.1 Experiment setup

Java programming language was used for implementing our algorithm in Eclipse and the simulations were performed using Net2Plan 0.5.0. All the experiments were carried out in a computer system with an Intel Core i5 processor and 8.00 GB of RAM. The service chains used in our experiments are generated randomly and they are at least 1 and at most 9 functions. We assumed that all the servers have equal capacity to host VNFs. There are 6 types of network functions where their required capacity is equal but the value varied in different scenarios. The number of demands varied in each scenario, ranging from 10 to 150 demands. The number of requested VNFs varied in each scenario, ranging from 30 to 270. The number of particles is 20, which is adopted from [30] and [31]. The value of acceleration factors is 2.05, which is adopted from [32].

Finally, the maximum number of iterations is 100, adopted from [33]. We work with three types of environments: (1) 32 servers, (2) 16 servers, and (3) 8 servers. Table 2 shows the details of the parameters used in this article.

**Table 2** Experimental parameters

| Item | Description/Value |
|---|---|
| Type of VNFs | Maximum 6 |
| Capacity of servers | Equal |
| Capacity of links | Equal |
| Number of VNFs in a service chain | Minimum 1 and maximum 9 |
| Number of particles | 20 |
| Maximum number of iterations | 100 |
| Required capacity of VNFs | Equal, depending on the scenario |
| Number of requested VNFs | Minimum 30 and maximum 270 |
| Number of demands | Minimum 10 and maximum 150 |

## 5.2 Simulation results

In this section, we discuss the results for experiments that were carried to evaluate the effectiveness of our algorithm. For each experiment, the experiment is carried 20 rounds and then the average value is considered.

### 5.2.1 Effect of service chain length

In this experiment, the network with 32 servers is considered, the demands are the same, and the number of demands is 30. The service chain length is varied in the range of 1 to 9 functions. Fig. 4 shows the effect of service chain length on execution time. The execution time is defined as the required time taken by the Random algorithm and our algorithm is run over the maximum number of iterations to find a near-optimal solution to solve the VNF placement and chaining problem. Meanwhile, the number of used servers, the average propagation delay of paths, and the average utilization of links is minimized.

According to Fig. 4, by increasing the number of VNFs in the service chain, the execution time increases as well. The increase is due to the more number of required VNF instances in the network, as each VNF has limited capacity to process. So, the required time to find servers to host more VNFs and selecting paths to meet the demands increases.

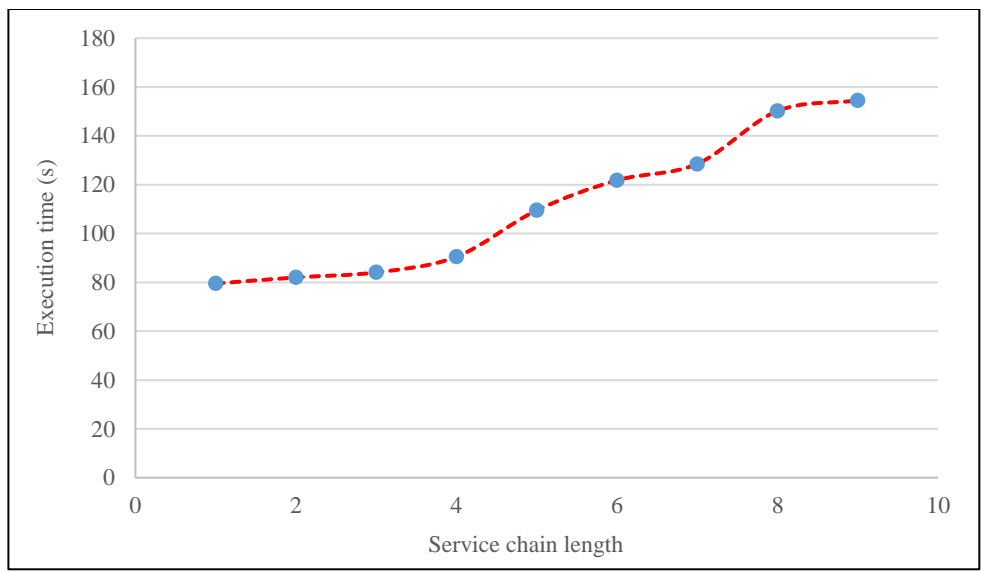

**Fig. 4** Effect of service chain length on execution time

For each demand in the network, a path should be determined. The path begins at the source server of that demand, traverses different links and servers, meets the required VNFs in a specific order, and ends at the destination server. Each path has a propagation delay that is computed as the sum of the propagation delays of all the links in that path. Fig. 5 shows that by increasing the number of VNFs in each service chain, the average propagation delay of all paths increases. This is because the number of required VNFs to meet increases. Also, considering the limited capacity of servers to host VNFs, the number of servers to host VNFs increases and leads to an increase in the length of the paths. Therefore, the average propagation delay of paths prolongs.

### 5.2.2 Effect of the number of servers

For the first part of this experiment, we collected data from 50 runs of 100 iterations of our algorithm, where the number of requested VNFs varied between 30 and 210 VNFs. We measured the execution time of our algorithm in three different networks. In this experiment, each demand contains at least 3 and at most 6 VNFs and has 6 types of VNFs. Fig. 6 shows the time needed to find near-optimal solutions in different networks when the number of requested VNFs increases.

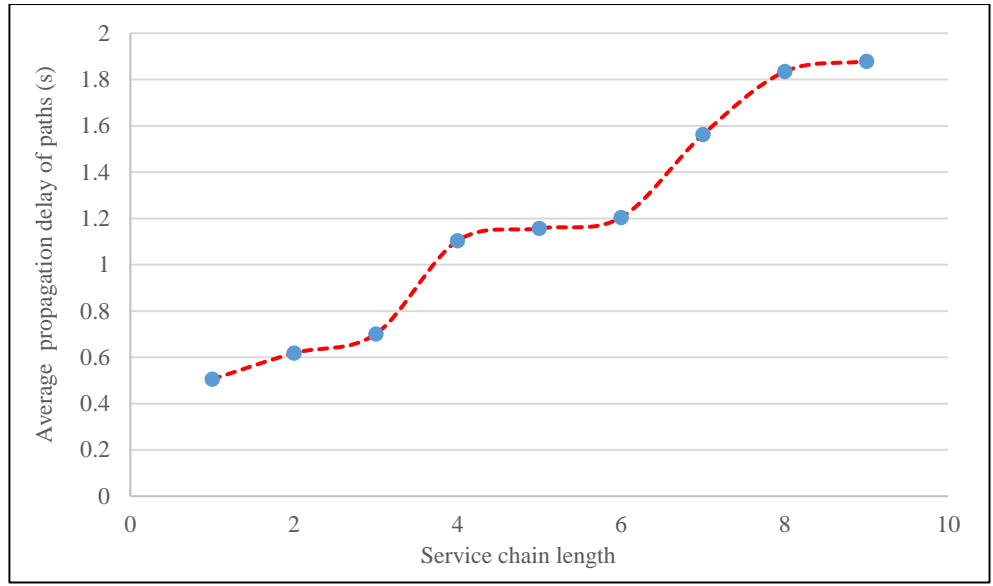

**Fig. 5** Effect of service chain length on the average propagation delay of paths

The first observation is that by increasing the number of servers in the network for a specific number of requested VNFs, the execution time increases. However, this increase is acceptable and our proposed algorithm scales reasonably with the size of the network. By increasing the number of servers, search space grows and the number of choices increases, so the execution time will increase.

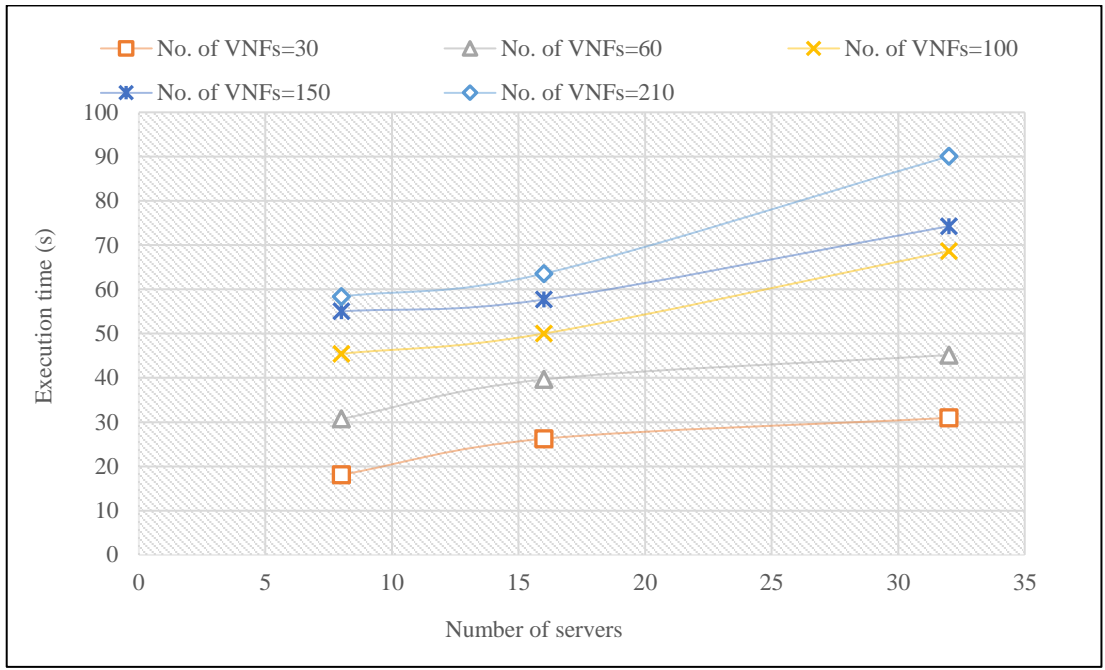

**Fig. 6** Effect of the number of servers on execution time with different number of requested VNFs

The second observation is that in a special network (e.g. 8 servers), by increasing the number of requested VNFs, the execution time increases. Increasing the number of requested VNFs in this scenario is due to the increase in the service chain length and an increase in the number of demands. By increasing the service chain length, the number of the required VNF instances in the network increases. Afterward, this increase causes extra time to find suitable servers to host VNFs. Moreover, increasing the number of demands causes increasing the number of used servers and the number of paths to meet the demands, because for each demand a path should be selected. Selecting more paths needs more time. This experiment shows the scalability of our algorithm for both the number of requested VNFs and the number of physical servers. Our results confirm that the execution time to find a near-optimal solution is more affected by the number of requested VNFs than the number of servers in the network.

For the next part of our experiment, we changed the required capacity for each VNF. We considered two scenarios for the required capacity of VNFs. In the first scenario, the required capacity of VNFs is equal and is 100 units while it is 160 units in the second scenario. When the required capacity of VNFs increases, a lower number of VNFs can be placed on a server and the number of required VNF instances decreases. The reason is that a VNF can serve serval demands and increasing the capacity of a VNF allows severing more demands. When the number of demands is small (i.e., 30 demands), as shown in Fig. 7, by increasing the capacity of VNFs, the execution time increases. This is because when the number of demands is small, the total traffic to be processed is low. Besides, since a VNF can serve several demands, the number of required VNFs is small. By considering the fixed and limited capacity of servers, when the required capacity of VNFs is high, the number of host servers increases because we need all the VNF types in the demands. Hence, the required time to select servers to host VNFs and selecting paths increases, suggesting an increase in the execution time. Therefore, it can be claimed that these results depend on the capacity of severs and the total traffic of demands.

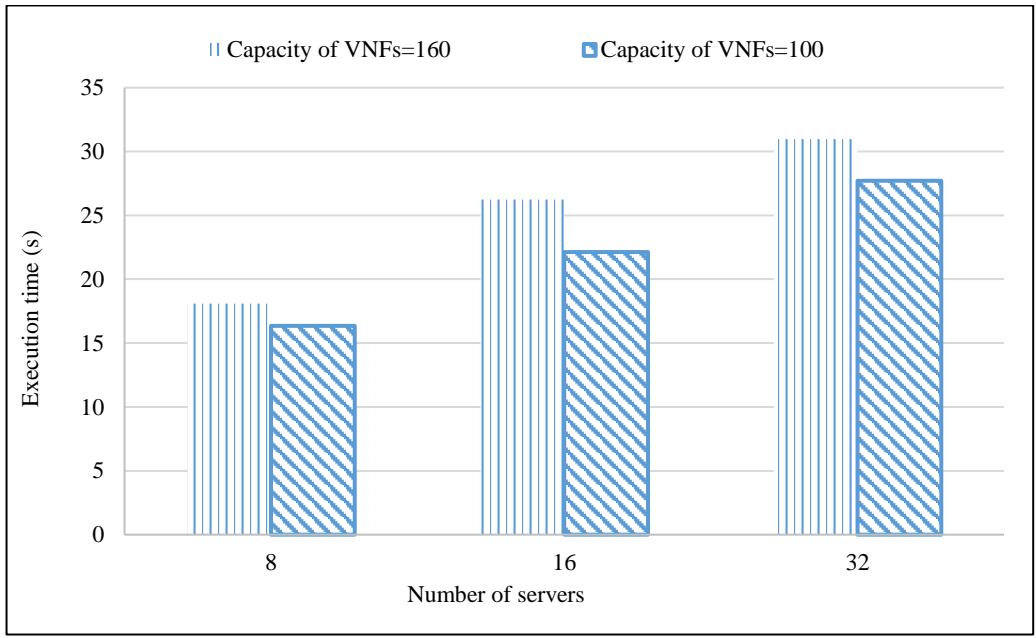

**Fig. 7** Execution time in different networks with a different required capacity of VNFs for 30 demands

When the number of demands increases (i.e., 150 demands), as shown in Fig. 8, the execution time decreases by increasing the capacity of VNFs. The reason is that when the number of demands is large, the total traffic to be processed by the VNFs is high, and thus more VNF instances are required. When the capacity of a VNF is high, it can serve more demands, and thus the number of required VNFs with high capacity decreases; meanwhile, the number of required VNFs with low capacity increases. This will cause an increase in the execution time for VNFs with low capacity. Nevertheless, this increase is small and negligible. Overall, it can be stated that these results will depend on the capacity of severs and the total traffic of demands.

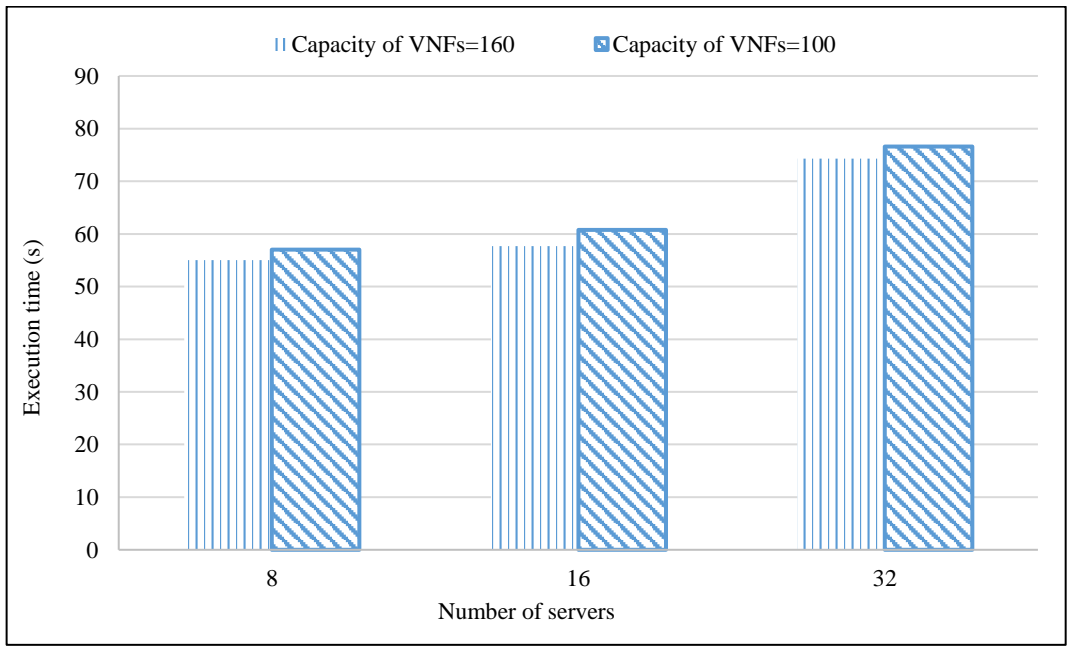

**Fig. 8** Execution time in different networks with a different required capacity of VNFs for 150 demands

### 5.2.3 Comparison of the proposed algorithm, BACON, and Greedy-K-NFV

In Figure 9, the proposed algorithm is compared with the Greedy-k-NFV algorithm [22], and the BACON model [41]. This comparison is performed under a small network of 32 servers and 4 types of VNFs that investigate the effect of locating and chain VNFs on route latency. It also evaluates the performance of our proposed algorithm. The maximum number of iterations of our proposed algorithm is 100 and the number of particles is 20. Also, as shown in Figure 9, our proposed algorithm produces the paths with the least latency. Then the BACON algorithm has less latency compared to the Greedy-k-NFV algorithm. Of course, our proposed algorithm not only minimizes route latency, but also minimizes the number of servers used and the bandwidth consumed by the links.

The database for comparing all three proposed algorithm methods, BACON, and Greedy-K-NFV is shown in Table 3.

**Table 3** network data set for comparison

| Set | Count |
|---|---|
| Available servers in network | 32 |
| VNF of type one | 2 |
| VNF of type two | 3 |
| VNF of type three | 2 |
| VNF of type four | 3 |

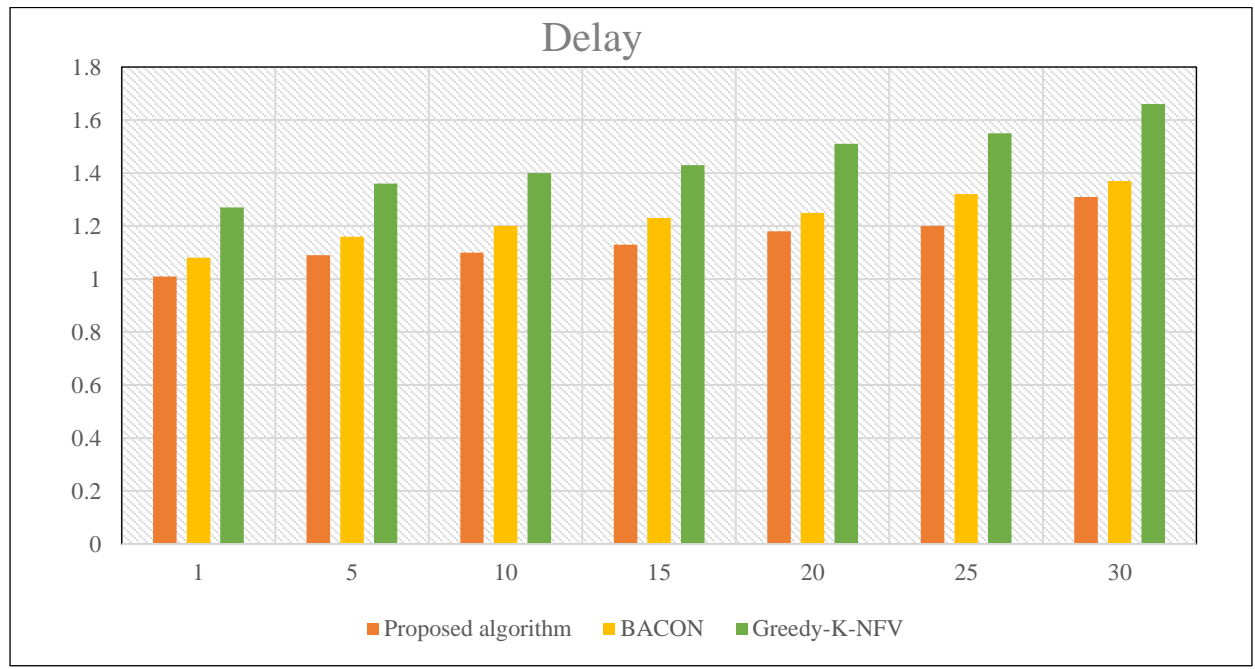

**Fig. 9** Comparison of the proposed algorithm with the BACON and Greedy-K-NFV methods.

## 5.2.4 Comparison of the proposed algorithm and Random algorithm

Fig. 10(a) shows the decrease in average utilization of links in our algorithm compared to the Random method in two networks for a different number of demands. In the Random approach, host servers are selected randomly such that the servers selected to host VNFs are far from each other. It will create longer paths to meet the demands and too many links to be traversed unnecessarily. As a result, traffic must traverse all the links in that path and it increases the average utilization of links. However, our algorithm decreases the average utilization of links considerably. The results of this study show that this algorithm compared to the random algorithm at least 53% and at most 81% decreases the average utilization of links. This algorithm performs much better in the larger networks because of the larger solution space.

Fig. 10(b) shows the number of used servers in our algorithm compared to the Random approach. In this article, we assume that if a server is going to host a VNF, it has to switch on, and then it is counted as a used server. Here, we want to minimize the number of servers used, so the ones that are switched on can be utilized efficiently. This will reduce power consumption and capital expenditure of the network.

In a random algorithm, servers are randomly selected to host virtual network functions and can be located anywhere on the network at any distance and number, and there is a high probability that the servers will not be fully utilized. The results show that compared to the random algorithm, our algorithm reduces the number of used servers.

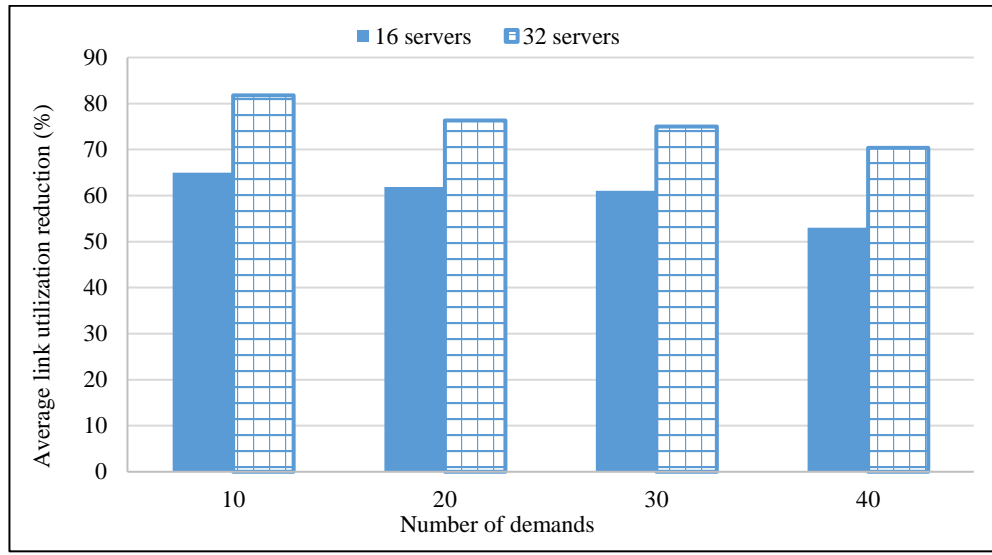

(a) Average link utilization reduction for different numbers of demands in two networks

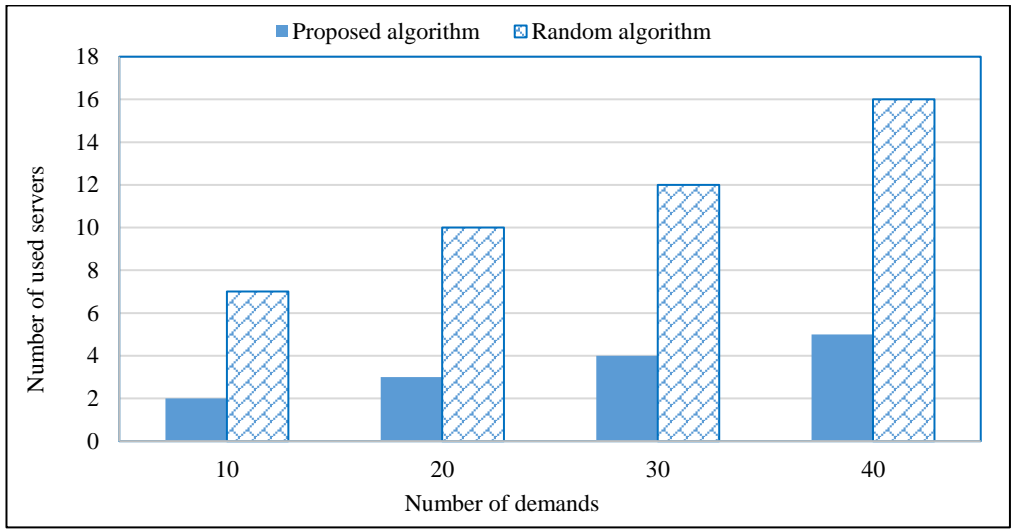

**(b)** the number of used servers,

**Fig. 10** Comparison of the proposed algorithm with the Random algorithm.

## 6 Conclusion

In this paper, an analysis of our proposed algorithm for the VNF placement and chaining problem is proposed. To this end, an optimization problem is presented to minimize the number of used servers, average utilization of links, and average propagation delay of paths. This algorithm can be used to determine the most suitable servers to host VNFs and the most suitable paths to steer traffic and meet the demands under network constraints. The results show that this algorithm can efficiently scale in networks composed of many required VNFs. Moreover, it was found that an increase in the number of servers and links and the execution time of our algorithm is more affected by the number of requested VNFs compared to the number of servers. The results also show that when there are many demands in the network, increasing the required capacity of VNFs can reduce the execution time compared to VNFs with less capacity and same conditions; however, depending on requested traffic of demands and capacity of servers, this increase can be small and, in some cases, negligible. Overall, the results of this study show that the proposed algorithm can reduce at least 53% and at most 81% of links utilizations compared to the Random algorithm, it can perform much better in larger networks, and it can reduce at most 76% of the average propagation delay of paths.